%% file: fide19.tex
\documentclass[submission,copyright,creativecommons]{eptcs}
\providecommand{\event}{F-IDE 2019} 
\usepackage{underscore}           
\usepackage{pdfpages}
\usepackage{amsmath}
\usepackage{color}
\usepackage{listings}
\usepackage{wrapfig}
\usepackage{amsfonts}
\usepackage{amssymb}
\usepackage{subfigure}
\usepackage{colortbl}
\usepackage{tabulary}
\usepackage{mathtools}

\title{Experience Report: Towards Moving Things with Types -- Helping Logistics Domain Experts to Control Cyber-Physical Systems with Type-Based Synthesis}
\author{Jan Bessai \qquad \qquad Moritz Roidl \qquad \qquad Anna Vasileva
\institute{Technical University of Dortmund\\ Dortmund, Germany}
\email{{jan.bessai, moritz.roidl, anna.vasileva}@tu-dortmund.de}
}
\def\titlerunning{Towards Moving Things with Types}
\def\authorrunning{J.Bessai, M. Roidl \& A. Vasileva}

\newcommand{\flwhalle}{logistics lab}

\normalsize

\begin{document}
\maketitle

\begin{abstract}
\input{abstract}
\end{abstract}
\input{introduction}

\input{cls}

\input{flwlab}
\input{extensions}
\input{lessons}


\bibliographystyle{eptcs}
\bibliography{bibliography}
\end{document}

%% file: abstract.tex
One of the ultimate goals of software engineering is to leave virtual spaces and move real things.
We take one step toward supporting users with this goal by connecting a type-based synthesis algorithm, Combinatory Logic Synthesizer ((CL)S), and its IDE to a logistics lab environment.
The environment is built and used by domain experts, who have little or no training in formal methods, and need to cope with large spaces of software, hardware and problem specific solution variability.
It consists of a number of Cyber-Physical Systems (CPS), including wheel-driven robots as well as flying drones, and it has laser-based support to visualize their possible movements.
Our work describes results on an experiment integrating the latter with (CL)S.
Possibilities and challenges of working in the domain of logistics and in cooperation with its experts are outlined.
Future research plans are presented and an invitation is made to join the effort of building better, formally understood, development tools for CPS-enabled industrial environments.

%% file: introduction.tex
\section{Introduction}
Logistics is the science (and sometimes art) of moving things.
Connections to computer science are immediate: paths must be planned, resources scheduled, object positions tracked, and automatization demands software systems to control physical actors.
Traditionally, logistics is an engineering discipline.
As such, it has an unspoken predisposition toward algorithms that use numeric methods for optimizing highly context-specific parameters.
Examples include Dijkstra's algorithm \cite{Dijkstra1959} for finding optimal paths and Kalman Filters \cite{kalman1960} for estimating the position of moving objects.
However, symbolic formal methods, which are the subject of large parts of computer science, are rarely transferred to the field.
Yet, it has all the potential for their application.
Systems scale from small packaging stations to country spanning interdependent logistical networks, while fine-tuned optimized numbers rarely scale at all. 
Hard- and software require consistent reasoning across daily evolving product lines of logistical equipment.
Pareto-optimal solution spaces can be vast or even infinite, in which case they need symbolic representations.
Finally, just to name a few of the potentials, formal guarantees can prevent catastrophic failures and thereby not only safe money but also lives, if systems are operated by humans.
This paper is part of a collaborative effort at the Technical University of Dortmund to identify and overcome reasons for the lack of transfer.
The authors are a mixed team of researchers in computer science (Bessai, Vasileva) and logistics (Roidl).
Starting point of the investigation is an experiment, which uses type-based synthesis to find ways through a labyrinth.
The scenario has been previously studied to illustrate progress on an IDE for the Combinatory Logic Synthesis, (CL)S, Framework \cite{Bessai.2018}.
It is described in Sec. \ref{sec:scenario}.
The problem of finding ways is sufficiently close to logistics to provide a starting point and to be practically evaluated in a logistics test lab, which is described in Sec. \ref{sec:lab}.
Conducting the experiment required extensions to existing software systems and it quickly became clear that new ways of visualizing solutions would be required to cater domain expert needs.
Extensions and results obtained this way are described in Sec. \ref{sec:extensions}.
They go beyond what was considered in the domain-independent design of the IDE previously developed for (CL)S.
Finally, some future plans and lessons learned from the ongoing cooperation are discussed in Sec. \ref{sec:lessons}.
We identify some key technical principles to make collaborations easier, and also focus on the social aspects, which can help to transfer knowledge.
The discussion of future plans includes an invitation to join our practical efforts, potentially by testing and improving other tools with us.

%% file: cls.tex
\section{Synthesizing Robot Paths with Types}\label{sec:scenario}

The implementation of (CL)S provides a framework for the automatic composition of software components from domain-specific repositories \cite{DBLP:conf/isola/BessaiDDMR16,DBLP:conf/isola/HeinemanBDR16, winkels2018automatic}. 
Automatic composition is performed by answering the type inhabitation problem of relativized Combinatory Logic with intersection types~\cite{JR13}
$\Gamma \vdash ? : \tau$.
That is, given a repository of typed combinators $\Gamma$, find all combinatory terms $M$ (inhabitants) of the goal type $\tau$.
The (CL)S Framework is implemented in Scala.
It is meant to be a tool for programmers as well as engineers with knowledge in programming who are not necessarily familiar with type theory.

In order to improve usability of the framework, we developed a web-based IDE \cite{ide, Bessai.2018}.
It is specifically focused to improve comprehensibility and traceability of the inhabitant search process.
The IDE provides a graphical overview of the inhabitants generated by the algorithm in form of hypergraphs \cite{Engelfriet,Kallat.2019}. 
We developed a step-wise build of the hypergraphs in order to explain the generation of solutions.
The web IDE also emits warnings if there are unused combinators or uninhabited types.
This exposes problematic specifications, which can be further analyzed in a perspective for presenting domain-specific repositories.

Additionally to the perspectives outlined above, which were previously discussed in \cite{Bessai.2018}, we developed features that focus on presenting local rather than global properties of solutions.
A perspective of the web IDE provides a list of inhabitants, s.t. a user can inspect each inhabitant and the corresponding hypergraph separately.
Intersection types allow combinators to have more than one type.
For instance, if we have a combinator:\\
$$down : (MovementPlan \to MovementPlan) \cap (Pos(0, 0) \to Pos(1, 0)) \cap (Pos(1, 2) \to Pos(2, 2))$$
it can be used to go from position $(0,0)$ to $(1, 0)$ and from position $(1,2)$ to $(2,2)$, while transforming a plan of movements to a plan of movements.
Inhabiting a type such as $MovementPlan \cap Pos(2, 2)$ requires selecting components form the type in a process called covering.
The IDE provides help to understand and debug this process for a given combinator and target, which is especially useful if combinator types are underspecified and lack required components.

\begin{wrapfigure}{l}{0.2\textwidth}
\begin{tabular}{r|p{0.3cm}|p{0.3cm}|p{0.3cm}|}
  & 0                 & 1                 & 2 \\\hline
0 &  				&         & \\\hline	
1 &                 &    \cellcolor{black}      & $\bullet$  \\\hline
2 &	        		&	 						&\\\hline
3 &    \cellcolor{black} &  $\bigstar$  & \cellcolor{black}\\\hline
\end{tabular}
\caption{Labyrinth example}
\label{fig:lab}
\end{wrapfigure}

We also developed a filtering function based on satisfiability modulo theories (SMT) \cite{Kallat.2019}.
In this way, additional constraints can be used to restrict the set of inhabitants and avoid trivial solutions such as $reverse(reverse(s))$ for some sequence $s$.
The IDE supports adding and removing some domain-independent structural constraints, which are helpful in many situations.

  

The synthesis of robot paths is based on the labyrinth example \cite{Bessai.2018}, an instance of which is shown in Fig.~\ref{fig:lab}.
The size of labyrinths, blocked paths (indicated by black boxes), and start (indicated by a black dot) as well as end positions (indicated by a star) are user-defined.
In Fig. \ref{fig:lab}, we have 3$\times$4 labyrinth with start position $Pos(1,2)$ and goal position $Pos(3,1)$. 
The repository $\Gamma$ includes combinators corresponding to the allowed move directions ($up, down, left$ and $right$), a combinator that provides the starting position as well as their type descriptions. 
The types in the repository represent all valid one-step moves.


%% file: flwlab.tex
\section{Logistics Research Lab}\label{sec:lab}
The research lab is designed for rapid prototyping of CPS \cite{concept40}. 
It is situated in an existing lightweight construction building that is similar to common industrial buildings used in logistics operations.  
It follows the basic concept of a highly flexible development testbed that is free of fixed or permanently installed equipment. 
The testbed is surrounded by several observation systems installed on the ceiling, at the walls, and within the floor.
It also includes a laser projection system, which is the most important component for this paper.
The experimentation space is 22 m long, 15 m wide, and up to 3.5-4 m high.
Eight laser projectors cover the full area and can project coloured vector graphics on the ground floor. 
The selection process favoured projectors with a high frame rate rather than accuracy. 
\begin{figure}[h!]
	\begin{minipage}[b]{0.5\textwidth}
		\includegraphics[width=8cm]{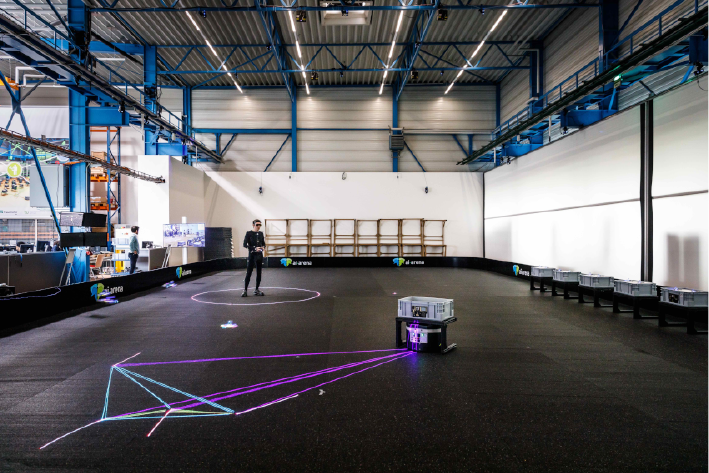}
		\caption{Logistics research lab overview}
		\label{fig:lab-overview}
	\end{minipage}
	\begin{minipage}[t]{0.5\textwidth}
		\hfill
        \includegraphics[width=7.5cm]{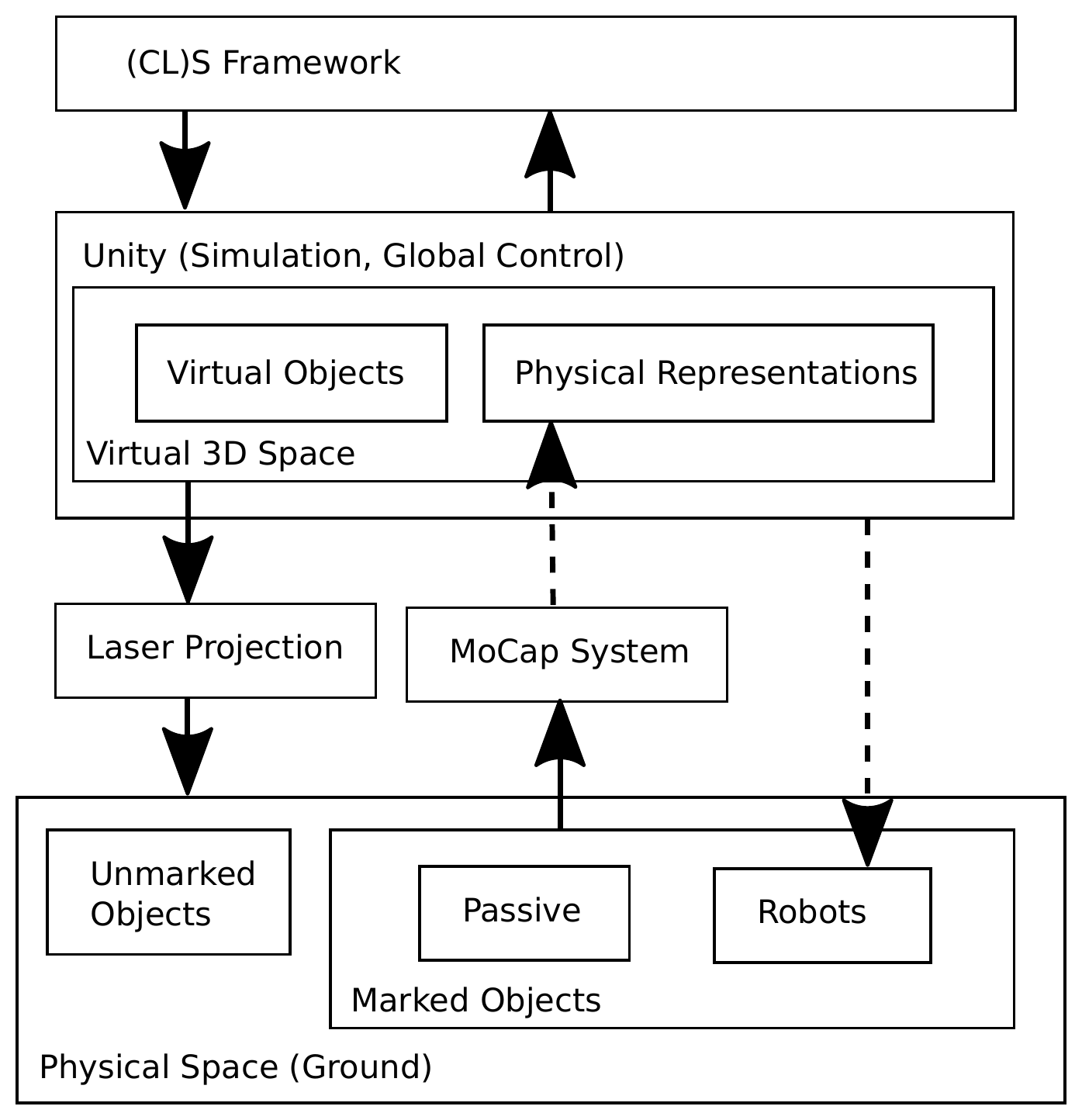}
		\caption{Logistics research lab architecture}
		\label{fig:lab-architecure}
	\end{minipage}
\end{figure}

Fig.~\ref{fig:lab-overview} shows a typical use of the system. 
A person, robot and box are tracked by the MoCap (Motion Capturing) system. 
The MoCap system consists of 40 infrared cameras by Vicon \cite{vicon}. 
The data stream is accessible over network to multiple clients and delivers position and rotation of tracked objects in three dimensions. 
In addition to simple physical objects, several marker-suits are available for tracking persons which are then used to generate data streams of complex skeletal models including individual body parts.
In the foreground, the laser system projects a visual representation of the current state of the steering algorithm of the robot. 
The projected circle around the person in the background represents the safety area which the robot is not allowed to enter. 
The box on the robot contains an embedded system (black square on the front) that communicates with the robot for transportation needs.



%% file: extensions.tex
\section{Experiment and Extensions}\label{sec:extensions}
The debugging facilities discussed in Sec.~\ref{sec:scenario} help with understanding types and to explore domain-independent technical aspects of the synthesis algorithm. 
However, a layer of interpretation is required to map them to any given concrete scenario.
Here diversity of training and mindsets really helped us to go forward.
While computer scientists working on formal methods are usually trained to aim for solutions with maximal feasible mathematical generality, engineers are focused on smallest viable solutions.
In the spirit of good engineering practices, we ventured to build a minimal prototype to demonstrate the following properties:\\
1. (CL)S is applied to a logistical setting within the demonstration capabilities of the \flwhalle.\\
2. Scenario-specific meaningful debugging is possible without consequences of damaged hardware.\\
3. Users interactively control the debugging process.\\
4. Hardware failures are separated from logical specification problems.\\
5. Existing technologies are used unchanged if possible, preferring adapters when necessary.\\
An explicit non-goal was to expect a realistic solution compliant with industry standards. 

\begin{figure}
	\begin{minipage}[t]{0.34\textwidth}
		\includegraphics[width=4cm]{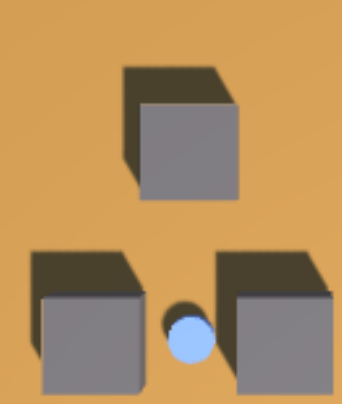}
		\caption{Unity 3D Representation}
		\label{fig:unity}
	\end{minipage}
	\begin{minipage}[t]{0.66\textwidth}
		\hfill
	\includegraphics[width=9cm]{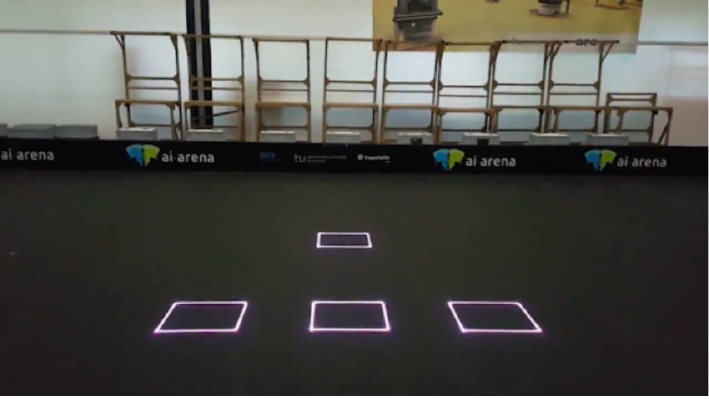}
	\caption{Laser Projection System Representation}
	\label{fig:labLaser}
	\end{minipage}
\end{figure}

The labyrinth example, incidentally, matched the first goal, which is why we decided to use it as a starting point.
Fig.~\ref{fig:lab-architecure} shows the architecture connecting it to the \flwhalle.
It contains four layers.
The bottom layer represents the real world.
It contains real entities (marked objects), which are robots and obstacles.
It is also augmented by laser projections of virtual objects (unmarked objects), e.g. start points and goals.
The second layer represents the fixed lab installation discussed above.
All marked objects in physical space are mirrored via live MoCap connection into a virtual environment that makes up the third layer.
It is realized using the Unity 3D~\cite{unity} game engine and capable of virtually representing physical states, simulating them when no connection to the lab is available, and controlling them by sending commands to the second layer.
Its virtual abstractions of physical state are mapped to component repositories for (CL)S, which is shown in the topmost layer.
Solutions synthesized by (CL)S are movement commands for robots and sent back to Unity, which either just simulates them in a 3D model or forwards them to the laboratory equipment.
In line with goals 2 and 4, we found it useful to virtualize robots, turning them into laser projections rather than immediately trying to control real hardware.
A 3D simulation in Unity and its laser projection are shown in figures~ \ref{fig:unity} and \ref{fig:labLaser}.
They show the end state of the labyrinth example (s. Section~\ref{sec:scenario}) where the robot has already reached the goal position (3, 1).
A video of the laser projected movements generated by (CL)S is available online \cite{video}.
Future development will fully exploit the possibility to update labyrinths based on real world obstacles detected by motion capturing, which would be in line with goal 3.

Connecting the different layers was a major challenge, because all of them are preexisting components implemented in different programming languages (Python and C++ for physical components, C\# for Unity and Scala for (CL)S).
The bottom layers were already connected by the ISO-standardized MQTT network protocol \cite{mqtt}, which made it natural to also use it for the communication with (CL)S.
Here, choosing a standard language to implement our tool was crucial, because we were able to use Eclipse Paho~\cite{paho}, an off-the-shelve implementation of MQTT, instead of going through the tedious error prone process of developing our own networking infrastructure.
Currently, all connections in Fig.\ref{fig:lab-architecure} are implemented using MQTT.
The dashed arrows have been implemented, but at the time of writing still need to be logically integrated into the mapping from Unity to (CL)S, which means labyrinths are currently specified in source code and robots are visualized with lasers.


%

%% file: lessons.tex
\section{Lessons Learned and Future Plans}\label{sec:lessons}
The experiment we presented is rather small and just a starting point, but already provided us with some technical and non-technical insights, which can be useful for other researchers willing to engage in interdisciplinary collaboration.
Technically, one of the most important aspects was to have a formal system with clear boundaries.
The (CL)S Framework is designed to collaborate with others by not insisting to share any of its platform specific requirements.
Instead, the experiment used a complete (even physical) separation with a light-weight MQTT network interface.
This allowed almost zero integration overhead and meant no software systems had to be rewritten.
The framework and system to control are both implemented in mainstream languages with preexisting library support for network communication.
We conjecture that some other platforms, e.g. with external domain specific languages, would have caused much higher development costs.
Being compatible with mainstream technology was key.
There was just no way we could have foreseen which networking libraries would be useful for our project.
Insisting on our own tech-stack would have caused the project to end before it started.
Also, on a technical level it quickly became clear, that domain-independent generic visualization methods were not sufficient.
While domain experts can appreciate that there is a formal model, for them it is just a vehicle to solve a real world problem.
Practically, this means specific tools are always preferred over abstract data.
For debugging and testing, users should be presented with a (graphical) language close to what they know.
Laser-based visualization and 3D-Modeling, which bridge the gap between virtual and physical systems, are perfect tools to provide a good, tangible user experience.
On a social level, small scale prototypes and specific solutions, that do not yet scale to a large system, are much more acceptable to engineers than generic solutions, if the smaller scale implies less effort to get real things moving.
This is very different from expert communities with more focus on mathematical theory, where generic solutions are expected and not filling in the details to get to a specific executable system is to be forgiven more easily, or even expected, because details are considered to be time-consuming and repetitive.
Another social lesson is to consider the training of the target audience.
Technicalities, such as obtaining the framework code from Git or executing it, imposed negligible effort, which is perhaps surprising for theory-minded people.
For our small scale experiment, conveying the purpose of synthesis, the meaning of intersection types and establishing a shared vocabulary took more time than the actual development.
While for us personal communication was the easiest way, this may not always be easily possible (e.g. if teams are situated in separate locations).
Future developments should take explaining the language of the tools into account.
We feel, that contact to other researchers is crucial to do this, because identifying conventions of language once they are established, is difficult.

In this spirit, extend our invitation to other researchers to collaborate on new experiments, possibly integrating more formal tools into the \flwhalle.
Our own upcoming next experiments will scale up the experiment to more realistic scenarios, where (CL)S synthesizes code computing paths instead of directly computing the paths.
Integrating some of the much more sophisticated existing path finding algorithms will allow the transition to a more fine granular view of the world than labyrinths with blocked paths.
Specifically it will enable the treatment of obstacles with non-rectangular shapes, non straight paths between points, speed and energy considerations during move and perhaps even reactive systems, where the position of obstacles can change (e.g. by having more robots).
In future CPS-based logistics systems, large numbers of autonomous and networked entities will arrange themselves ad-hoc in temporary constellations to provide logistics services in coordination with humans. 
The development of these heterogeneous systems presents a challenge to engineers with their complexity of interacting hard- and software in industrial environments.
It is our opinion that user-friendly formally understood development tools will be crucial to have any chance of facing this challenge.